\documentclass[showpacs,twocolumn,prb]{revtex4-1}
\usepackage{amsfonts}
\usepackage{graphicx}
\usepackage{amsmath}
\usepackage{amssymb}

\begin{document}

\title{Impact of Dresselhaus vs. Rashba spin-orbit coupling on the Holstein polaron}
\author{Zhou Li$^1$, L. Covaci$^2$, and F. Marsiglio$^1$}
\affiliation{$^1$ Department of Physics, University of Alberta, Edmonton, Alberta, Canada, T6G~2J1 \\
$^2$Departement Fysica, Universiteit Antwerpen, Groenenborgerlaan 171, B-2020 Antwerpen, Belgium}

\begin{abstract}
We utilize an exact variational numerical procedure to calculate the
ground state properties of a polaron in the presence of Rashba and
linear Dresselhaus spin-orbit coupling. We find that when the linear
Dresselhaus spin-orbit coupling approaches the Rashba spin-orbit
coupling, the Van-Hove singularity in the density of states will be
shifted away from the bottom of the band and finally disappear when
the two spin-orbit couplings are tuned to be equal. The effective
mass will be suppressed; the trend will become more significant for
low phonon frequency. The presence of two dominant spin-orbit
couplings will make it possible to tune the effective mass with more
varied observables.
\end{abstract}

\date{\today}
\maketitle





\section{Introduction}

One of the end goals in condensed matter physics is to achieve a sufficient understanding of
materials fabrication and design so as to `tailor-engineer' specific desired properties into a material.
Arguably $pn$-junctions long ago represented some of the first steps in this direction; nowadays, heterostructures\cite{bimberg99}
and mesoscopic geometries\cite{murayama01} represent further progress towards this goal.


In the field of \emph{spintronics}, where the spin degree of freedom
is specifically exploited for potential applications,\cite{wolf01,awschalom09}
spin-orbit coupling \cite{winkler03} plays a critical role
because control of spin will require coupling to the orbital motion.
Spin orbit coupling, as described by Rashba\cite{rashba60} and Dresselhaus,\cite{dresselhaus} is expected to be prominent in two dimensional
systems that lack inversion symmetry, including surface states. These different kinds of coupling are in principle
independently controlled.\cite{maiti11,meier07}

The coexistence of Rashba and Dresselhaus spin-orbit coupling has now been
realized in both semiconductor quantum wells\cite{meier07,awschalom09} and more
recently in a neutral atomic Bose-Einstein condensate.\cite{lin11} When
the Rashba and (linear) Dresselhaus spin-orbit coupling strengths are tuned to
be equal, SU(2) symmetry is predicted to be recovered and the
persistent spin helix state will emerge.\cite{bernevig06,awschalom09,lin11}
This symmetry is expected to be robust against spin-independent
scattering but is broken by the cubic Dresselhaus spin-orbit
coupling and other spin-dependent scattering which may be tuned to
be negligible.\cite{awschalom09}

While we focus on the spin-orbit interaction, other interactions are present.
In particular, the electron-phonon interaction will be present and may be
strong in semiconductor heterostructures. Moreover, optical lattices\cite{bloch08} with
cold polar molecules may be able to realize a tuneable Holstein model.\cite{herrera11}
The primary purpose of this work is to investigate the impact of electron-phonon coupling (as
modelled by the Holstein model\cite{holstein59}) on the properties of the spin-orbit coupled system.
We will utilize a tight-binding framework; previously it was noted that in the presence of Rashba spin-orbit coupling
the vicinity of a van Hove singularity near the bottom of the electron band\cite{cappelluti07,covaci09,li11} had a significant
impact on the polaronic properties of an electron; with additional (linear) Dresselhaus spin-orbit coupling the van Hove singularity shifts
well away from the band bottom, as the two spin-orbit couplings acquire equal strength.
As we will illustrate below, the presence of two separately tunable spin-orbit couplings will result in
significant controllability of the electron effective mass.


\section{Model and methodologies}

We use a tight-binding model with dimensionless Holstein electron-phonon coupling of strength $g$, and with linear
Rashba ($V_R$) and Dresselhaus ($V_D$) spin-orbit coupling:

\begin{eqnarray}
H &=&-t\sum_{<i,j>,s=\uparrow \downarrow }(c_{i,s}^{\dagger
}c_{j,s}+c_{j,s}^{\dagger }c_{i,s})  \notag \\
&&+i\sum_{j,\alpha ,\beta }(c_{j,\alpha }^{\dagger }\hat{V}_{1}c_{j+\hat{y},\beta
}-c_{j,\alpha }^{\dagger }\hat{V}_{2}c_{j+\hat{x},\beta }-h.c.)  \notag \\
&&-g\omega _{E}\sum_{i,s=\uparrow \downarrow }c_{i,s}^{\dagger
}c_{i,s}(a_{i}+a_{i}^{\dagger })+\omega _{E}\sum_{i}a_{i}^{\dagger }a_{i}
\end{eqnarray}
where $c_{i,s}^{\dagger }(c_{i,s})$
creates (annihilates) an electron at site $i$ with spin index $s$, and $a_{i}^{\dagger
}$ ($a_{i}$) creates (annihilates) a phonon at
site $i$. The operators $\hat{V}_j$, $j=1,2$ are written in terms of the spin-orbit coupling strengths and
the Pauli matrices as $\hat{V}_{1} = V_{R}\hat{\sigma}_{x} - V_{D}\hat{\sigma}_{y}$,
and $\hat{V}_{2} = V_{R}\hat{\sigma}_{y} - V_{D}\hat{\sigma}_{x}$,
The sum over $i$ is over all sites in the lattice, whereas
$<i,j>$ signifies that only nearest neighbour hopping is included.
Other parameters in the problem are the phonon frequency, $\omega_E$, and the hopping parameter $t$, which hereafter is set equal
to unity.

\begin{figure}[tp]
\begin{center}
\includegraphics[height=2.0in,width=2.0in,angle=-90]{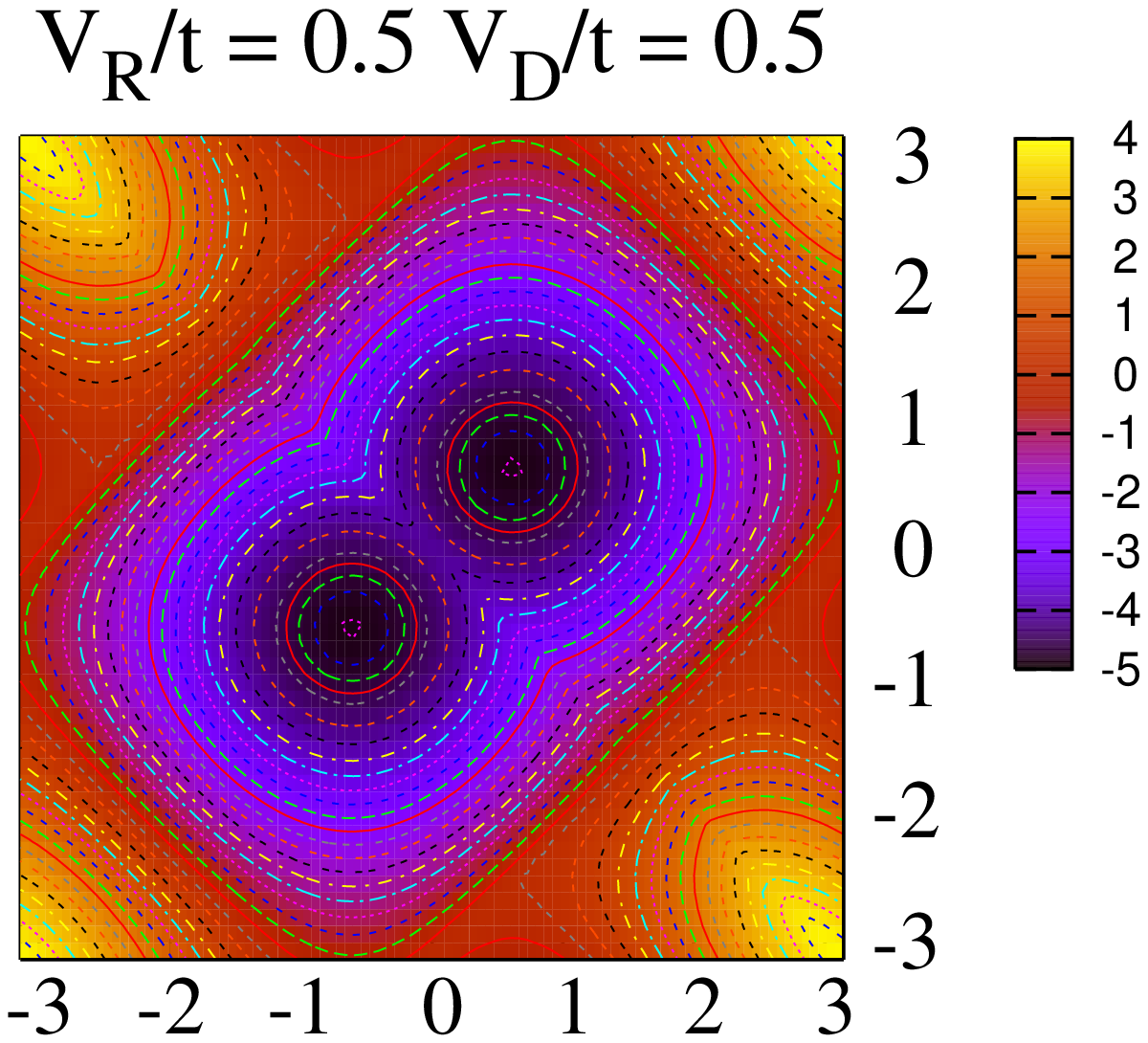} %
\includegraphics[height=2.0in,width=2.0in,angle=-90]{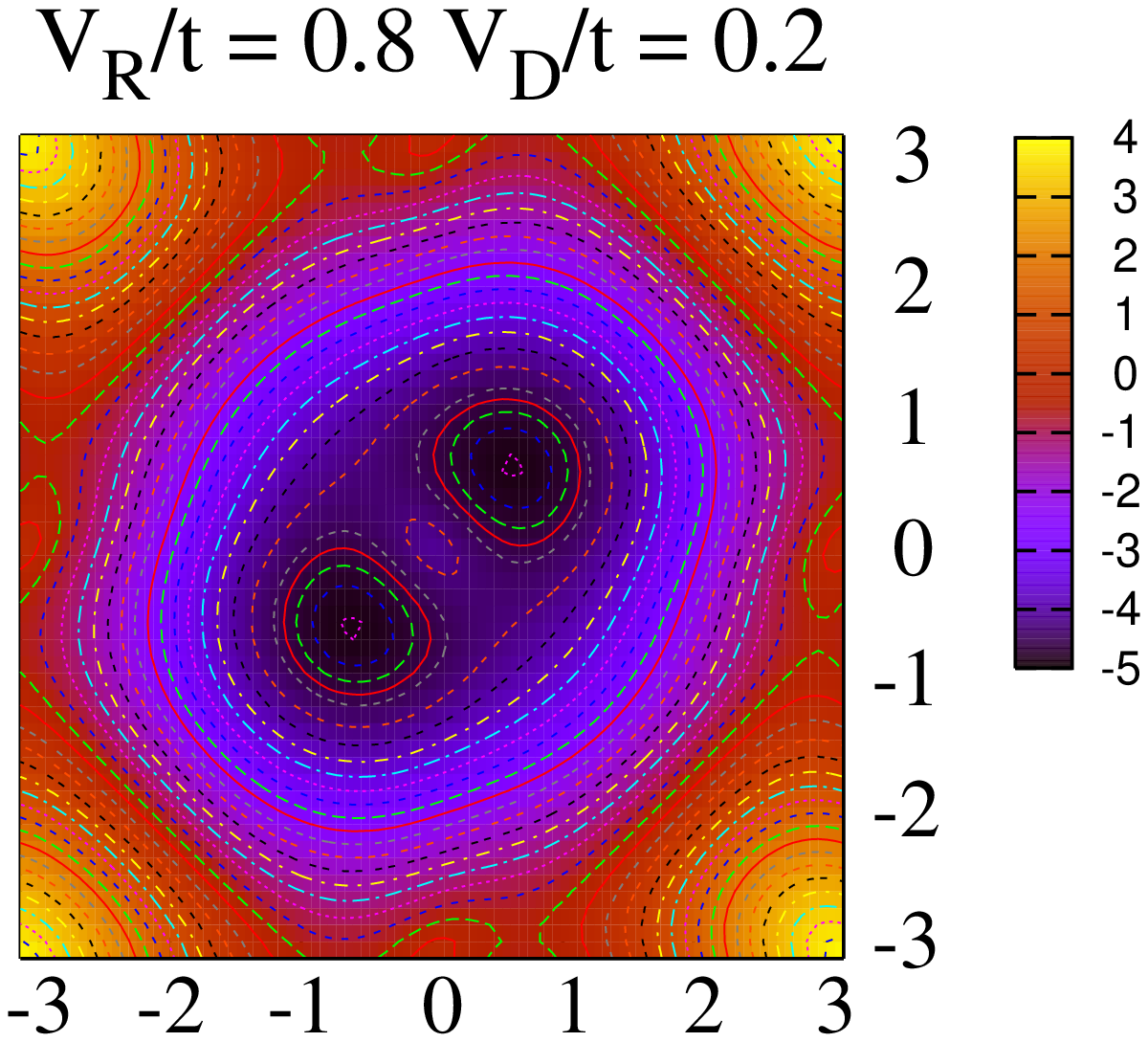} %
\includegraphics[height=2.0in,width=2.0in,angle=-90]{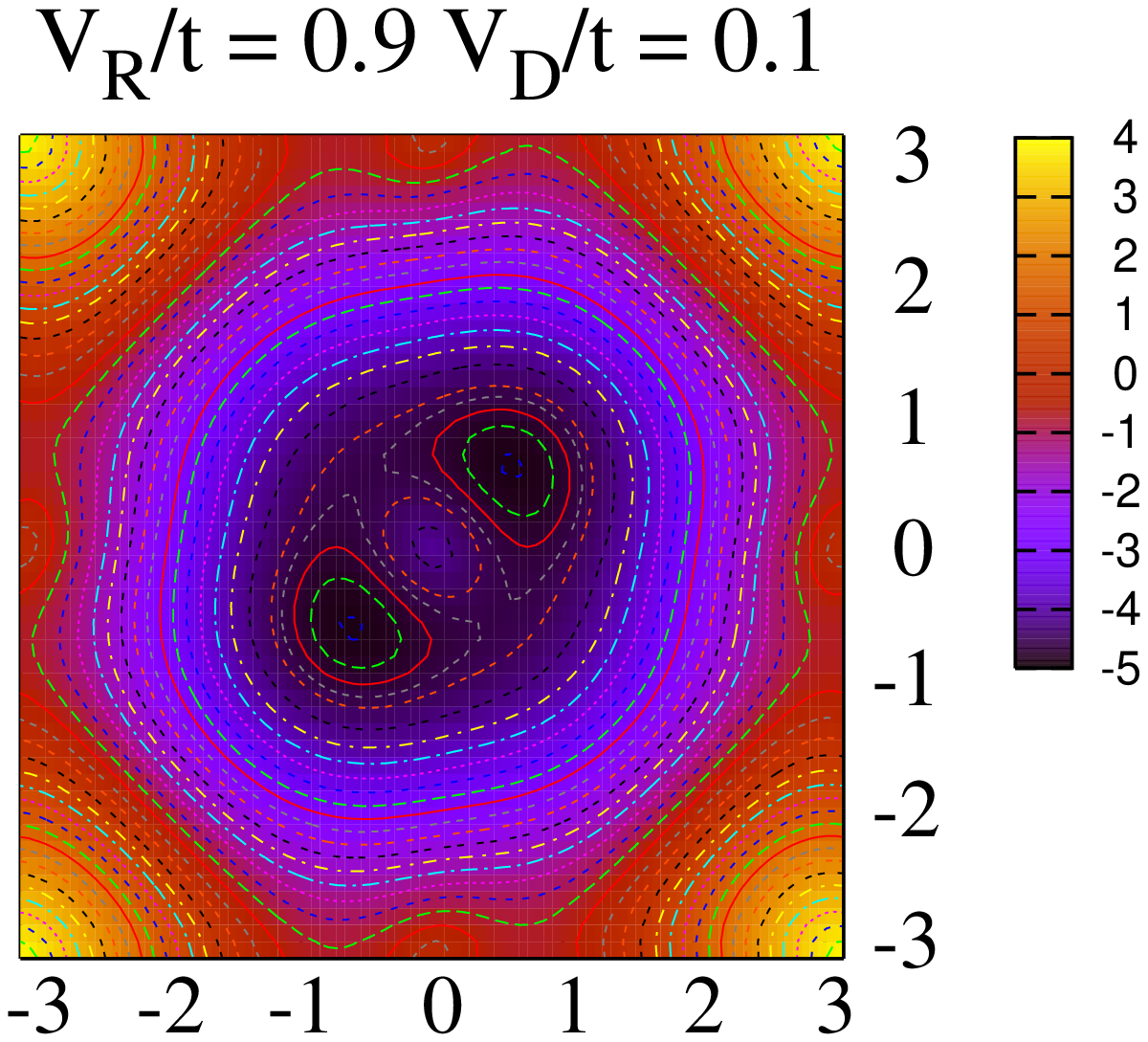} %
\includegraphics[height=2.0in,width=2.0in,angle=-90]{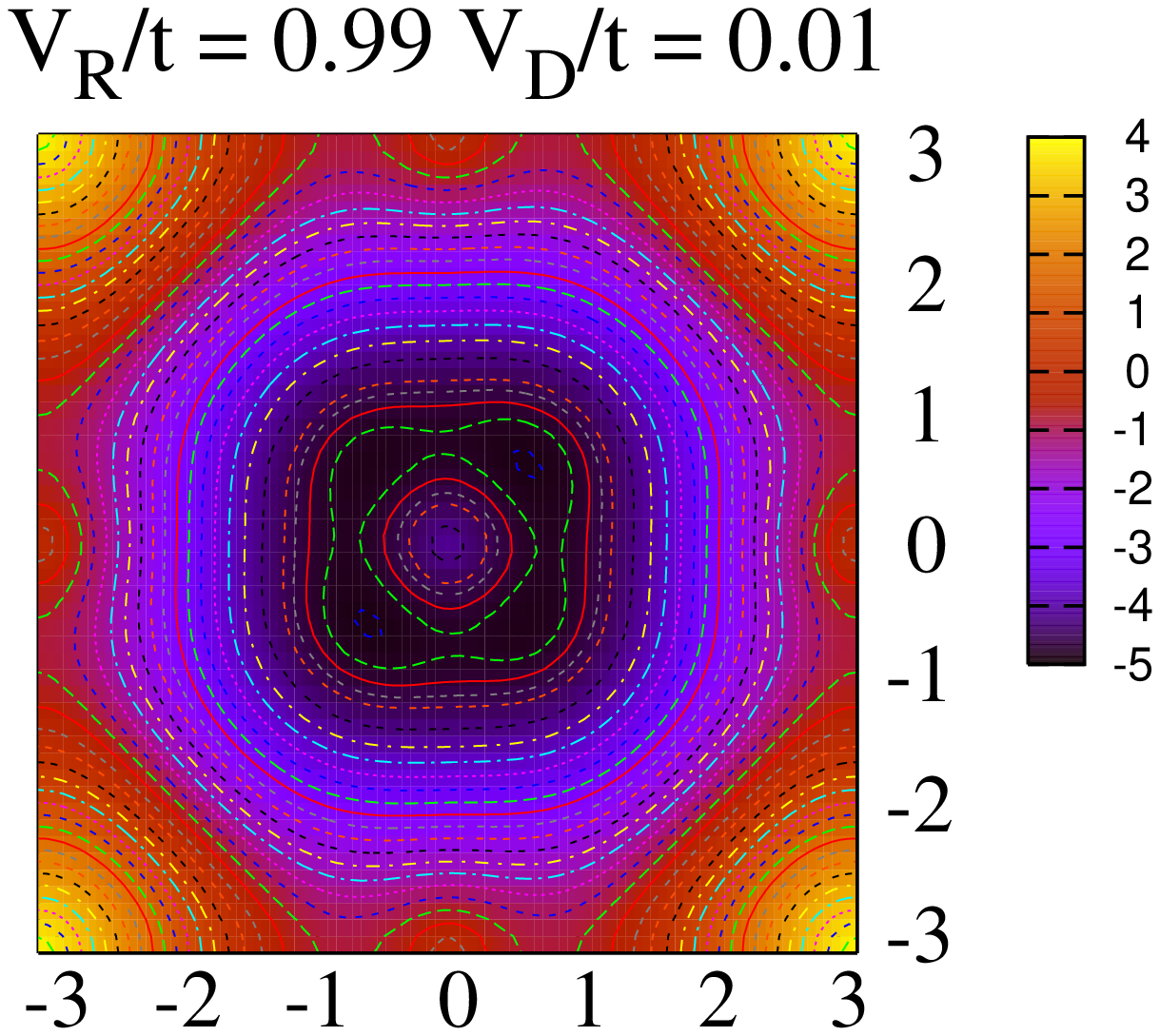}
\end{center}
\caption{Contour plots for the bare energy bands with Rashba-Dresselhaus spin-orbit coupling, for different values of
$V_{R}$ and $V_{D}$ while the sum is kept constant: $V_{R}+V_{D}=t$ for these cases.
(a) $V_{R}=V_{D}=0.5t$, (b) $V_{R}=0.8t,V_{D}=0.2t$,
(c) $V_{R}=0.9t,V_{D}=0.1t$, and (d) $V_{R}=0.99t,V_{D}=0.01t$. Note the clear progression from a two-fold degenerate
ground state to a four-fold degenerate one.}
\label{fig1}
\end{figure}
Without the electron-phonon interaction the electronic structure is readily obtained
by diagonalizing the Hamiltonian in momentum space. With the definitions
\begin{eqnarray}
S_1 &\equiv& V_{R}\sin (k_{y}) + V_{D}\sin (k_{x}), \nonumber \\
S_2 &\equiv& V_{R}\sin (k_{x}) + V_{D}\sin (k_{y}),
\label{defns}
\end{eqnarray}
we obtain the eigenvalues
\begin{equation}
\varepsilon _{k,\pm }=-2t[\cos (k_{x})+\cos (k_{y})]\pm 2 \sqrt{S_1^2 + S_2^2}
\label{eigenvalues}
\end{equation}
and eigenvectors
\begin{equation}
\Psi _{k \pm }=\frac{1}{\sqrt{2}}\left[ c_{k\uparrow }^{\dagger }\pm \frac{S_1 - i S_2}{\sqrt{S_1^2 + S_2^2}} c_{k\downarrow }^{\dagger }\right] |0\rangle  .
\label{eigenstates}
\end{equation}
The ground state energy is
\begin{equation}
E_0 = -4t\sqrt{1 + (V_R + V_D)^2/(2t^2)}.
\label{groundstate_energy}
\end{equation}
Without loss of generality we can consider only  $V_R \ge 0$ and $V_D \ge 0$.
Either Rashba and Dresselhaus spin-orbit coupling independently behave in the same manner, and give rise to
a four-fold degenerate ground state with wave vectors,  $(k_x,k_y) = (\pm \arctan (\frac{V_{R}}{\sqrt{2}t}),\pm \arctan (\frac{V_{R}}{\sqrt{2}t})$, ($V_D = 0$),
and similarly for $V_D \ne 0$ and $V_R = 0$. With both couplings non-zero, however, the degeneracy becomes two-fold, with
the ground state wave vectors,
\begin{equation}
(k_{x0},k_{y0}) = \pm (k_0,k_0); \phantom{a} {\rm where} \phantom{a} k_0 = {\rm tan}^{-1} (\frac{V_{R} + V_{D}}{\sqrt{2}t}).
\label{gs_wave_vectors}
\end{equation}
It is clear that the sum of the coupling strengths replaces the strength of either in these expressions,
so that henceforth in most plots we will vary one of the spin-orbit interaction strengths while maintaining their sum to be fixed.
Similarly, the effective mass, taken along the diagonal, is
\begin{equation}
\frac{m_{\mathrm{SO}}}{m_{0}}=\frac{1}{\sqrt{%
1+(V_{R}+V_{D})^{2}/(2t^{2})}},  \label{eff_mass_so}
\end{equation}
where $m_{0}\equiv 1/(2t)$ (lattice spacing, $a \equiv 1$, and $\hbar \equiv 1$) is the bare mass in the absence of
spin-orbit interaction, and $m_{\mathrm{SO}}$ is the effective mass
due solely to the spin-orbit interaction. As detailed in the Appendix, the effective mass becomes isotropic when the Rashba and Dresselhaus
spin-orbit coupling strengths are equal.


\begin{figure}[tp]
\begin{center}
\includegraphics[height=3.5in,width=3.5in,angle=0]{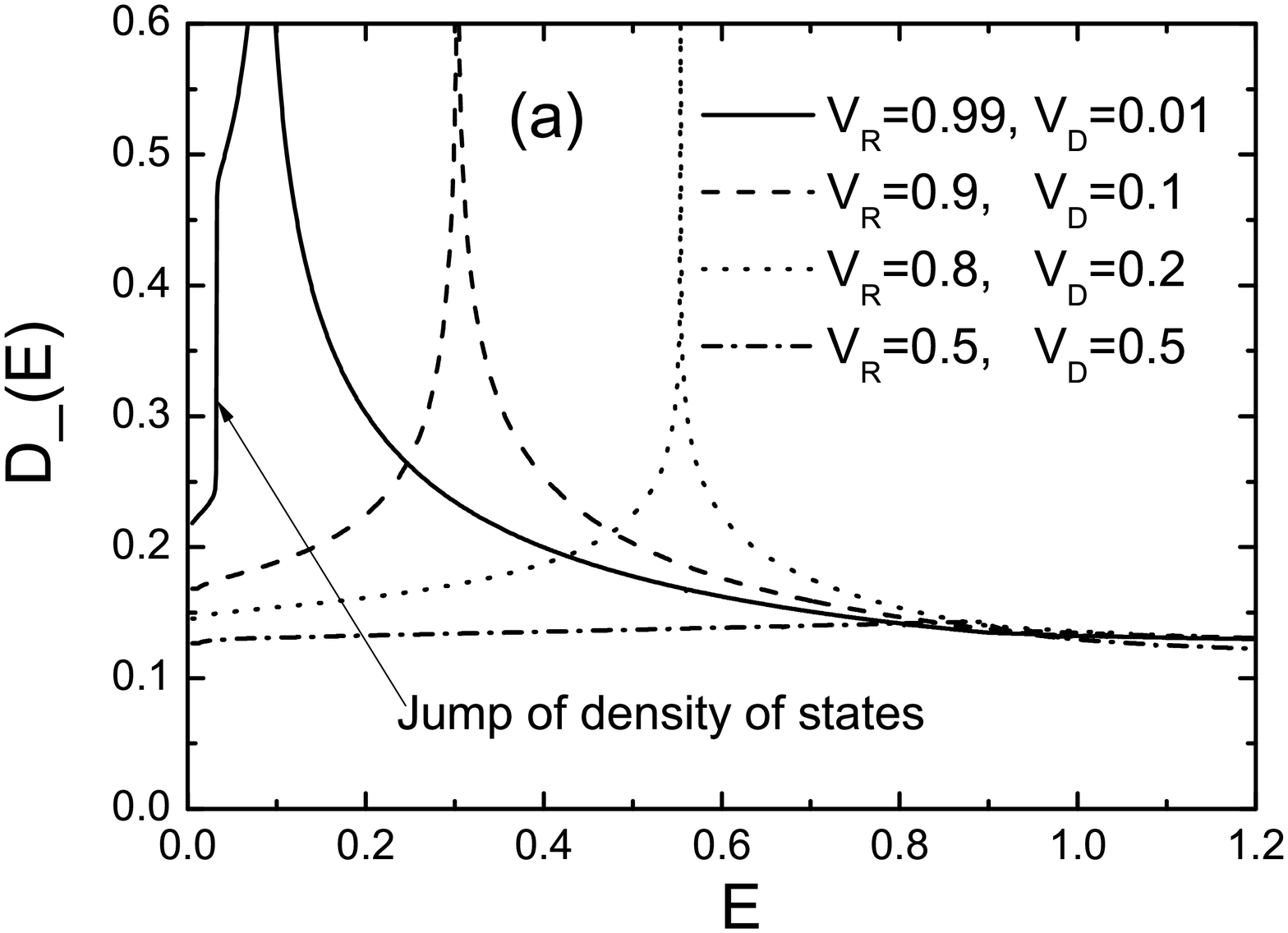} %
\includegraphics[height=3.5in,width=3.5in,angle=0]{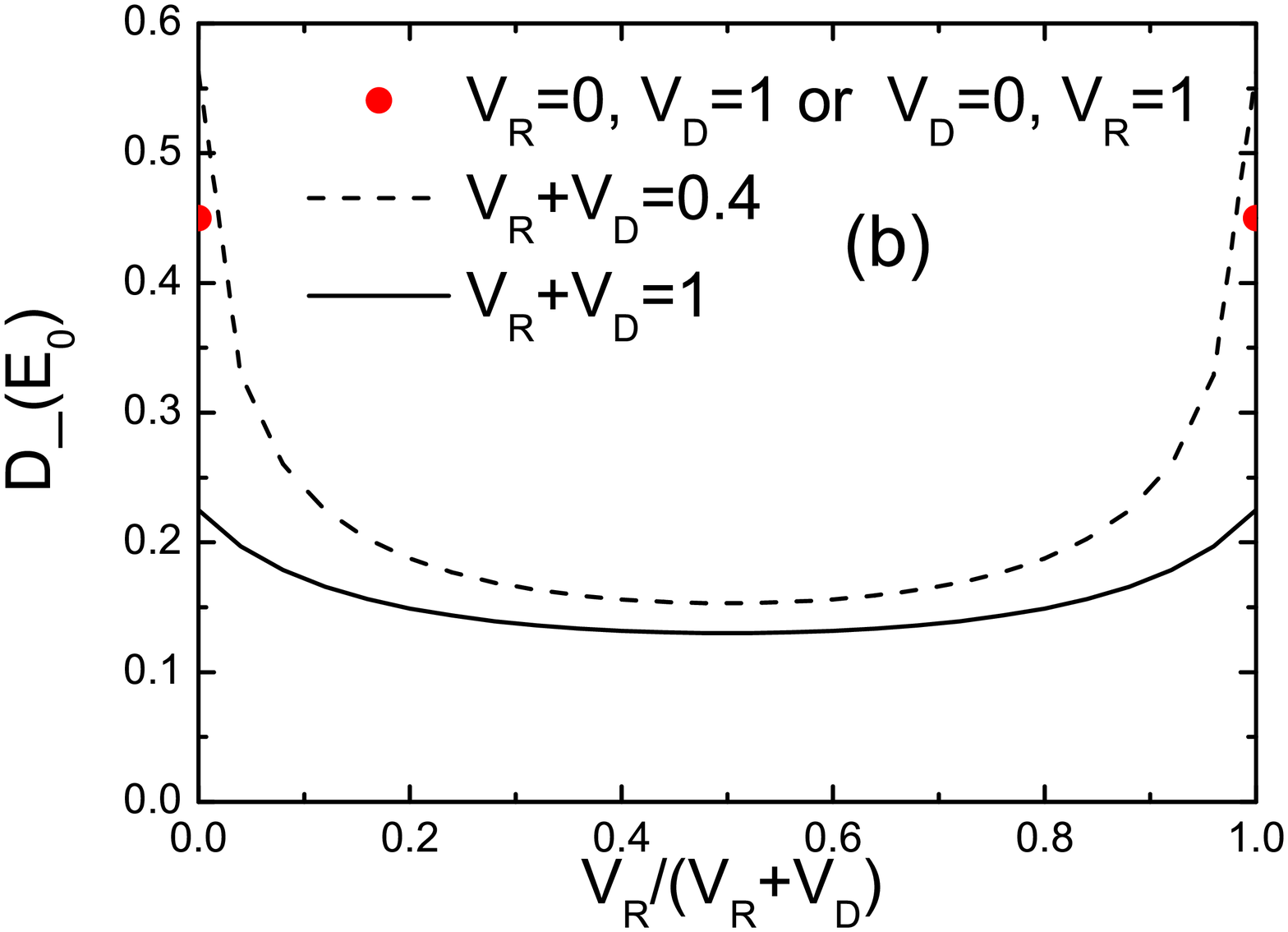}
\end{center}
\caption{(a)The non-interacting density of states $D_{-}(E)$ near the bottom of the
band for four values of the spin-orbit coupling strengths: $(V_{R},V_D)/t = (0.5,0.5)$ (dot-dashed curve), $(0.8,0.2)$ (dotted curve),
$(0.9,0.1)$ (dashed curve), and $(0.99,0.01)$ (solid curve).
Note that for equal coupling strengths there is no van Hove singularity at low energies.
(b) The value of the density of states at the bottom of the band (ground state) as a function of
$V_{D}$ (while the total coupling strength, $V_{R} + V_{D}$, is held constant. The value of the density of states
achieves a minimum value when $V_{R} = V_{D}$. For $V_{R}=0$ or $V_{D}=0$ there is a discontinuity,
caused by the transition from a
doubly degenerate ground state to a four-fold degenerate ground state.}
\label{fig2}
\end{figure}

The non-interacting electron density of states (DOS) is defined for each
band, as
\begin{equation}
D_s(\epsilon) = \sum_k \delta(\epsilon - \epsilon_{ks})  \label{dos}
\end{equation}
with $s =\pm 1$.

In Fig.\ref{fig2}(a) we show the low energy DOS for various values of the
spin-orbit coupling strengths, $V_{R}$  and $V_{D}$, while keeping their sum constant;
the low energy van Hove singularity disappears for
$V_{R}=V_{D}$. Note that only
$D_{-}(\epsilon)$ is shown, as the upper band, with DOS $D_+(\epsilon)$, exists only at higher energies. Furthermore,
information concerning the upper band can always be obtained through the
symmetry
\begin{equation}
D_{+}(\epsilon)=D_{-}(-\epsilon).  \label{sym}
\end{equation}%
In Fig.\ref{fig2}(b) we show the value of the density of states at the bottom of the
band vs. $V_D$; as derived in the Appendix, the DOS value at the minimum energy is given by
\begin{equation}
D_{-}(E_{0})=\frac{1}{2\pi t}\frac{1}{\sqrt{1+\frac{(V_{R}+V_{D})^{2}
}{2t^{2}}-\frac{(V_{R}-V_{D})^{2}}{(V_{R}+V_{D})^{2}}}}.
\end{equation}%
Note that when the coupling strengths are equal, the density of states has a
minimum. Also note that when one kind of spin-orbit coupling vanishes, e.g. $%
V_{R}=0$, or $V_{D}=0$, there will be a discontinuity for the
density of states (the density of states jumps to twice its value). This is
caused by a transition from a doubly degenerate ground state to a four-fold
degenerate ground state. This discontinuity will also appear for $%
V_{D}\simeq 0$ or $V_{R}\simeq 0$ near the bottom of the band as can
be seen from Fig.\ref{fig2}(a) for $V_{R}=0.99,V_{D}=0.01.$

\section{Results with the electron-phonon interaction}

As the electron phonon interaction is turned on, the ground state
energy (effective mass) will decrease (increase) due to polaron
effects. To study the polaron problem numerically, we adopt the
variational method outlined by Trugman and
coworkers,\cite{trugman90,bonca99} which is a controlled numerical technique to determine polaron
properties in the thermodynamic limit exactly. This method was
recently further developed\cite{li10,alvermann10} to study the polaron
problem near the adiabatic limit with Rashba spin-orbit coupling.\cite{li11}
This case was also studied in Ref. [\onlinecite{covaci09}] using the Momentum Average Approximation.\cite{berciu06}

\begin{figure}[tp]
\begin{center}
\includegraphics[height=3.5in,width=3.5in,angle=0]{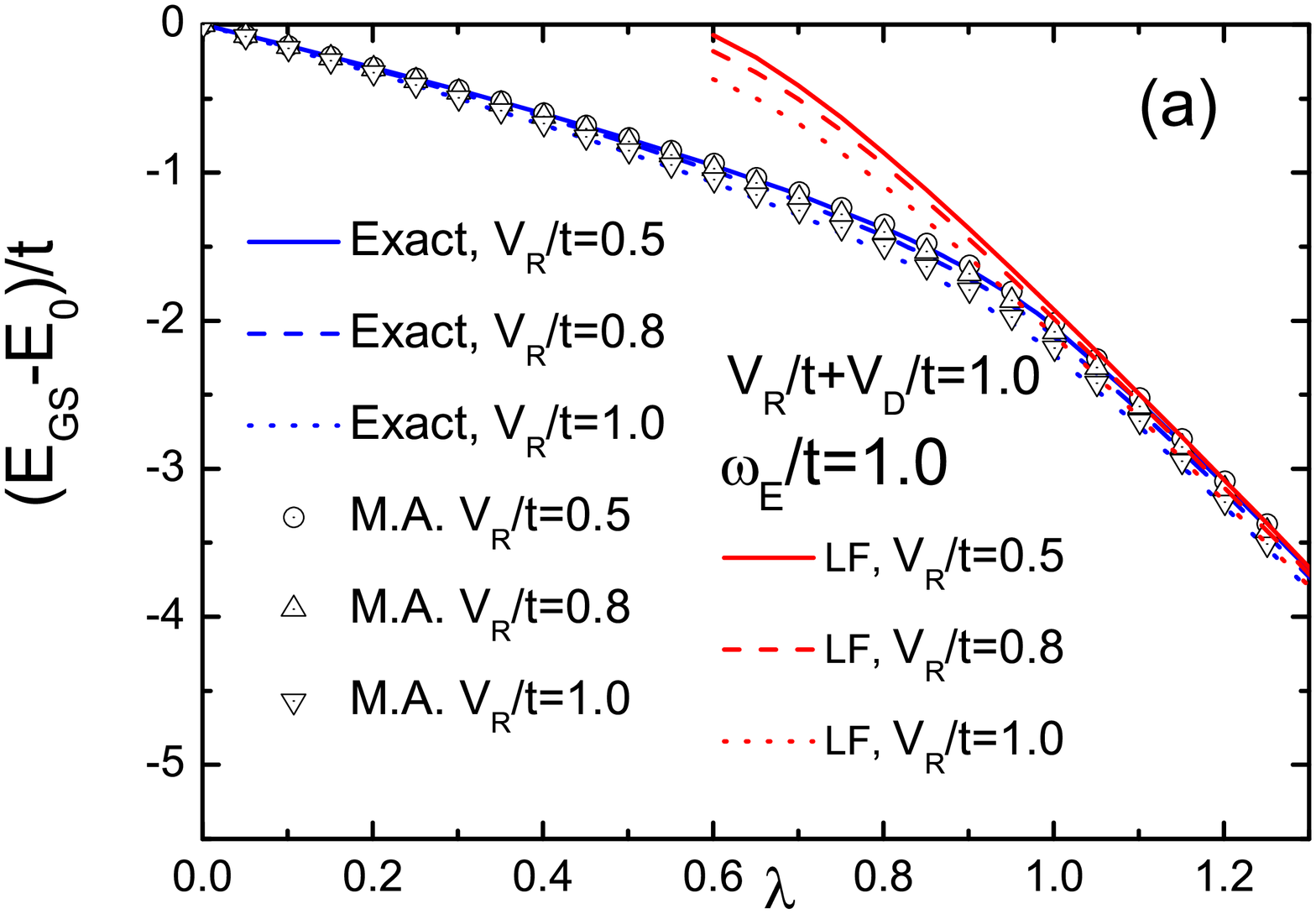} %
\includegraphics[height=3.5in,width=3.5in,angle=0]{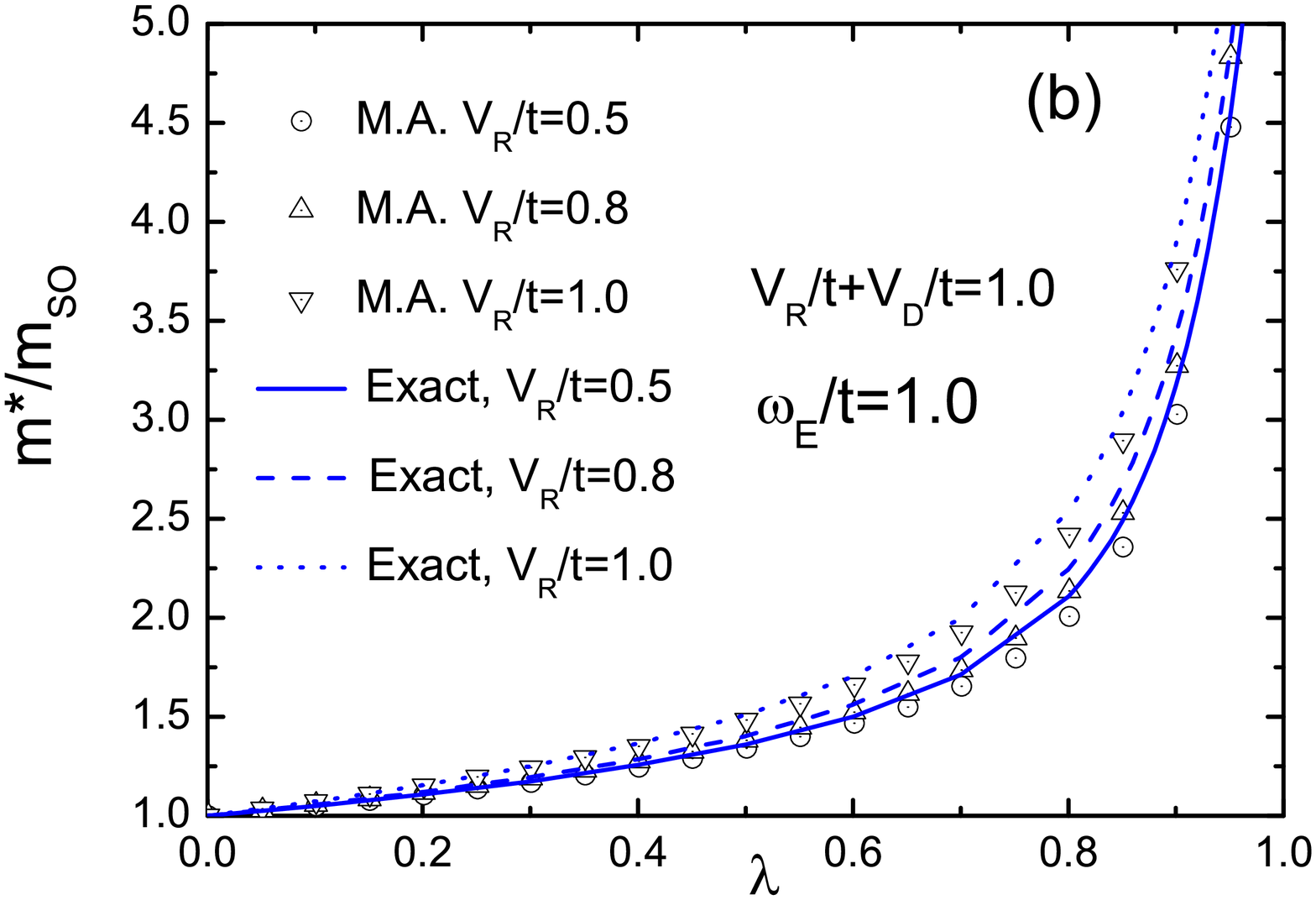}
\end{center}
\caption{(a) Ground state energy difference $E_{GS}-E_{0}$ vs. $\protect%
\lambda $ for $V_{R}/t=0.5,0.8,1.0$ and $\protect\omega
_{E}/t=1.0$ while the total coupling strength is kept fixed: $V_{R}+V_{D}=t$. Exact numerical results
are compared with those from the Momentum Average (MA) method. Agreement is excellent.
Strong coupling results are also plotted (in red) by utilizing the
Lang-Firsov (LF) strong coupling approximation. Agreement in the strong coupling regime ($\lambda \ge 1$) is
excellent. (b) Effective mass $m^{\ast}/m_{SO}$ vs. $%
\protect\lambda $. MA results are plotted (symbols) with the exact numerical results, and again, agreement is excellent.
In both (a) and (b) the polaronic effects are minimized for $V_{R} = V_{D}$. } \label{fig3}
\end{figure}

\begin{figure}[tp]
\begin{center}
\includegraphics[height=3.5in,width=3.5in,angle=0]{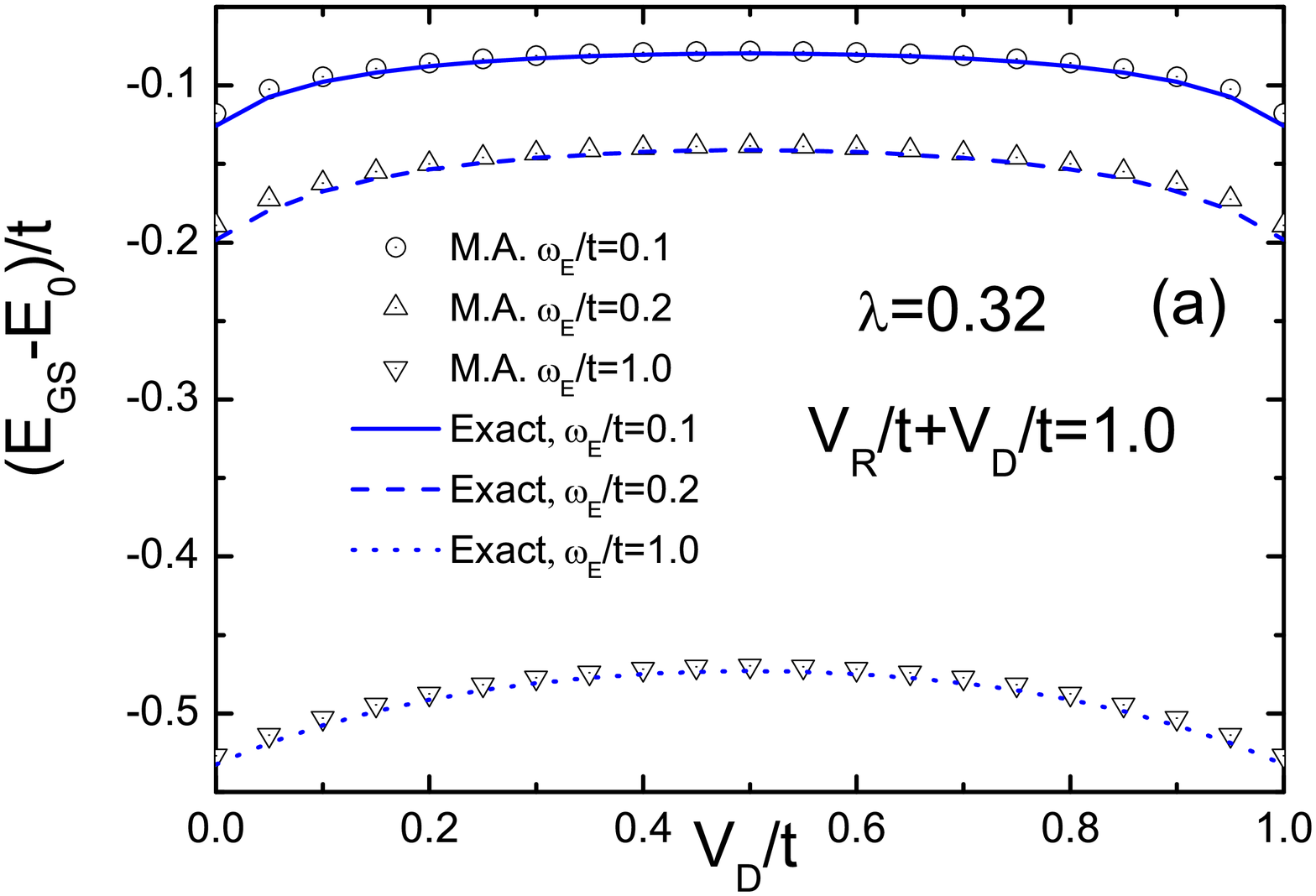} %
\includegraphics[height=3.5in,width=3.5in,angle=0]{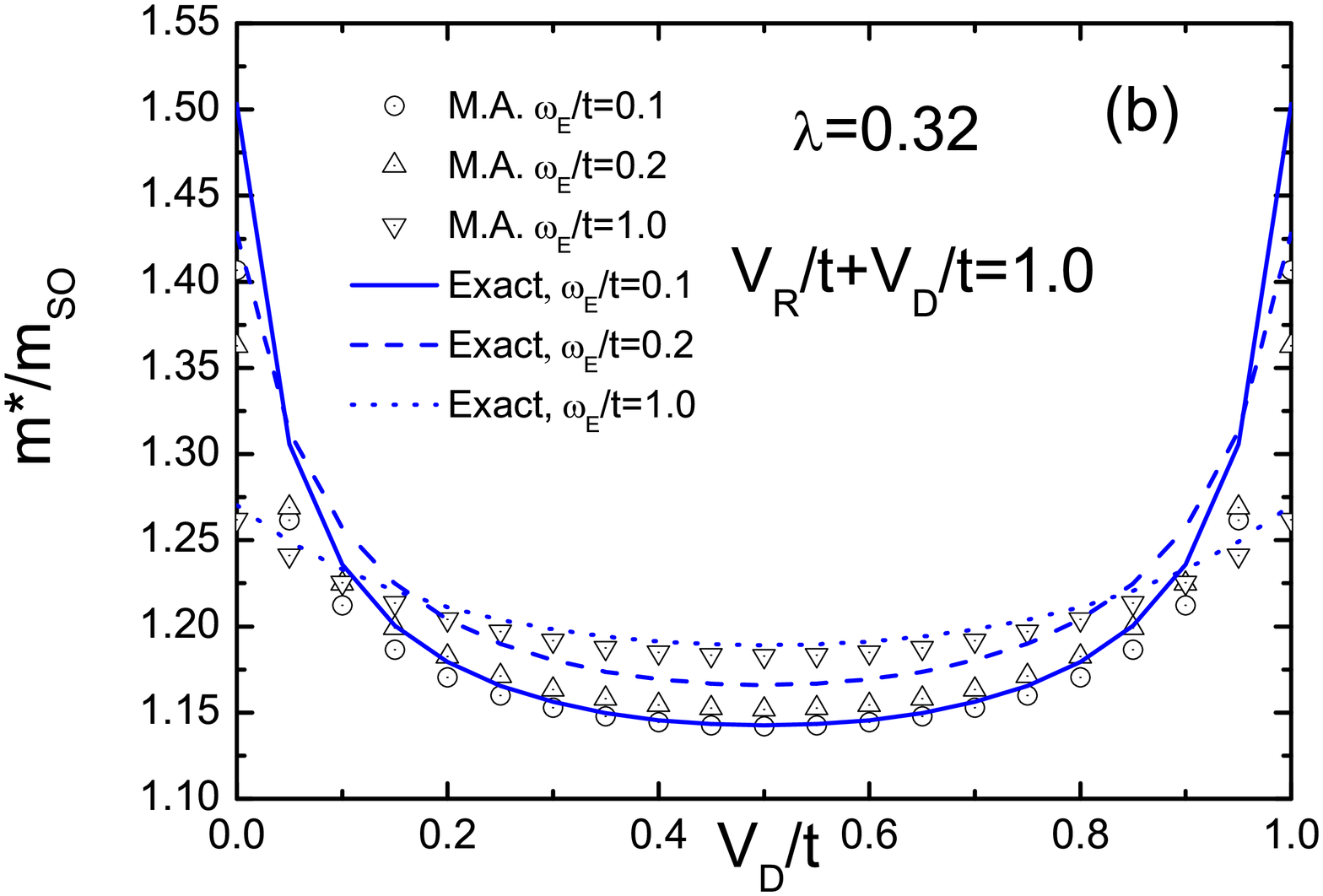}

\end{center}
\caption{(a) Ground state energy $E_{GS}-E_{0}$ as a function of
spin orbit
coupling $V_{D}/t$ for $\protect\omega_{E}/t=0.1,0.2,1.0$ with weak electron phonon coupling,
$\protect\lambda =0.32$, and moderate spin-orbit coupling, $V_{R}+V_{D}=t$. (b) Effective mass
$m^{\ast }/m_{SO}$ as a function of spin orbit coupling $V_{D}/t$
for the same parameters. MA results are again compared with the exact numerical results, and are
reasonably accurate for these parameters.}
\label{fig4}
\end{figure}

In Fig.~\ref{fig3}, we show the ground state energy and the effective
mass correction as a function of the electron phonon coupling $\lambda \equiv 2g^2 \omega_E/(4\pi t)$,\cite{li10}
for various spin-orbit coupling strengths, but with the sum fixed: $V_{R}+V_{D}=t$.
These are compared with the results from the Rashba-Holstein model
with $V_{D}=0$. Here the phonon frequency is set to be $\omega
_{E}/t=1.0$, which is the typical value used in Ref.[\onlinecite{covaci09}], and for each value of $V_{R}$,
the ground state energy is compared to the corresponding result for $
\lambda =0$. The numerical results are compared with results from the MA method
and from Lang-Firsov strong coupling theory\cite{lang63,marsiglio95} (see Appendix). In Fig.~\ref{fig3}(a), the
ground state energy crosses over smoothly (at around $\lambda \approx 0.8$)
from the delocalized electron regime to the small polaron regime. In the
whole regime, the ground state energy is shifted up slightly as the Dresselhaus spin-orbit coupling, $V_D$, is
increased in lieu of the Rashba spin-orbit coupling. We show results for $V_D \le V_R$, as the complementary regime
is completely symmetric. The MA results agree very well with the exact results and the Lang-Firsov strong coupling results
agree well in the $\lambda \ge 1$ regime. Similarly, weak coupling perturbation theory\cite{li11} agrees with the exact
results for $\lambda \le 1$ (not shown).  Fig.~\ref{fig3}(b) shows the effective mass as a function of coupling strength; it decreases
slightly, for a given value of $\lambda$, by increasing $V_D$ in lieu of $V_R$.

All these results are plotted as a function of the electron phonon coupling strength, $\lambda$, as defined above; this definition requires
the value of the electron density of states at the bottom of the band, and we have elected to use, for any value of spin-orbit coupling, the
value $1/(4 \pi t$ appropriate to {\it no} spin-orbit coupling. If the actual DOS appropriate to the value of spin-orbit coupling were used in the
definition of $\lambda$, then the effective mass, for example, would vary even more with varying $V_D$ vs. $V_R$ (see Fig.~2(b)). Moreover,
this variation would be more pronounced for lower values of $\omega_E$.

In Fig.~\ref{fig4}, we show results for the ground state energy and effective mass
for different values of the Einstein phonon frequency, $\omega_E$; MA results are also shown
for comparison. In these plots the electron phonon coupling strength is kept fixed and $V_D$ is varied while maintaining
the total spin-orbit coupling constant. The ground state energy has a maximum when the two spin-orbit coupling strengths,
$V_{D}$ and $V_{R}$, are tuned to be equal; similarly,
the effective mass has a minimum when the two are equal. As the phonon frequency is reduced the minimum in the effective mass
becomes more pronounced. The MA results track the exact results, and, as found previously,\cite{li11} are slightly less accurate as the
phonon frequency becomes much lower than the hopping matrix element, $t$.

\section{Summary}

Linear spin-orbit coupling can arise in two varieties; taken on their own, they are
essentially equivalent, and their impact on a single electron, even in the presence of electron phonon interactions,
will be identical. However, with the ability to tune either coupling constant, in both solid state and cold atom experiments,
one can probe the degree of Dresselhaus vs. Rashba spin-orbit coupling through the impact on polaronic properties.
The primary effect of this variation is the electron density of states, where the van Hove singularity can be moved as a function
of chemical potential (i.e. doping) through tuning of the spin-orbit parameters. These conclusions are based on exact methods (the so-called Trugman method),
and are not subject to approximations. These results have been further corroborated and understood through the Momentum Average approximation,
and through weak and strong coupling perturbation theory. The effect is expected to be experimentally relevant since in typical materials with large spin-orbit couplings the phonon
frequency is small when compared to the bandwidth, $\omega_E/t \ll 1$.

\begin{acknowledgments}

This work was supported in part by the Natural Sciences and
Engineering Research Council of Canada (NSERC), by ICORE (Alberta),
by the Flemish Science Foundation (FWO-Vl) and by the Canadian
Institute for Advanced Research (CIfAR).
\end{acknowledgments}

\appendix

\section{Density of States and effective mass}

Expanding $\varepsilon _{k,-}$ around the minimum energy $E_{0}$, by
defining $k_{x}^{\prime }=k_{x}\pm \arctan (\frac{V_{R}+V_{D}}{\sqrt{%
2}t}),k_{y}^{\prime }=k_{y}\pm \arctan (\frac{V_{R}+V_{D}}{\sqrt{2}t}%
),$ we have
\begin{equation}
\varepsilon _{k,-} = E_{0} + \tilde{t}_1 \bigl\{ k_{x}^{\prime 2}+k_{y}^{\prime 2}\bigr\} \pm
\tilde{t}_2 k_x^\prime k_y^\prime,
\end{equation}
where
\begin{equation}
\tilde{t}_1 = t \Biggl\{\frac{1+\frac{(V_{R}+V_{D})^{2}}{2t^{2}}-\frac{%
(V_{R}-V_{D})^{2}}{2(V_{R}+V_{D})^{2}} }{\sqrt{1+(V_{R}+V_{D})^{2}/(2t^{2})}} \Biggr\},
\end{equation}
and
\begin{equation}
\tilde{t}_2 = t \Biggl\{\frac{\frac{(V_{R}-V_{D})^{2}}{(V_{R}+V_{D})^{2}}}{\sqrt{1+(V_{R}+V_{D})^{2}/(2t^{2})}} \Biggr\},
\end{equation}
Note that, with generic spin-orbit coupling, the effective mass is in general anisotropic, but when $V_D = V_R$,
it becomes isotropic.

To calculate the density of states at the bottom of the band, from the
definition, we have
\begin{equation}
D_{-}(E_{0}+\delta E)=\frac{1}{4\pi ^{2}}\int_{-\pi }^{\pi }dk_{x}\int_{-\pi
}^{\pi }dk_{y}\delta (E_{0}+\delta E -\varepsilon _{k,-}),
\end{equation}%
where $\delta E$ is a small amount of energy above the bottom of the band, $%
E_{0}$. Around the two energy minimum points there are two small regions
which will contribute to this integral. We choose one of them (and then
multiply our result by a factor of 2), then use the definitions of $k^{\prime }$
above instead of $k$, and introduce a small cutoff $k_{c}$, which is the radius
of a small circle around $k_{\min }$. Thus the integral becomes
\begin{eqnarray}
&&D_{-}(E_{0}+\delta E)=2\times \frac{1}{4\pi ^{2}}\int_{0}^{k_{c}}k^{\prime
}dk^{\prime }\int_{-\pi }^{\pi }d\theta  \notag \\
&& \delta \biggl[ \delta E-  \bigl\{ \tilde{t}_1 +
\frac{1}{2}\tilde{t}_2 \sin 2\theta \bigr\} k^{\prime 2}
\biggr] \notag \\
&=&\frac{1}{2\pi t}\frac{1}{\sqrt{1+\frac{(V_{R}+V_{D})^{2}}{2t^{2}}-%
\frac{(V_{R}-V_{D})^{2}}{(V_{R}+V_{D})^{2}}}}
\end{eqnarray}%
In the weak electron-phonon coupling regime, perturbation theory can be
applied to evaluate the effective mass; the self energy to first order in $%
\lambda $ is given by
\begin{equation}
\Sigma _{\mathrm{weak}}(\omega +i\delta )=\pi \lambda t\omega _{E}\sum_{%
\mathbf{k,}s=\pm }\frac{1}{\omega +i\delta -\omega _{E}-\varepsilon _{k,s}}.
\label{weak_a}
\end{equation}%
The effective mass can be obtained through the derivative of the self energy
\begin{equation}
\frac{m_{\mathrm{weak}}^{\ast }}{m_{SO}}=1-\frac{\partial }{\partial \omega }%
\Sigma _{\mathrm{weak}}(\omega +i\delta )|_{\omega =E_{0}}.  \label{weak_b}
\end{equation}%
By inserting the expansion of $\varepsilon _{k,-}$ around the minimum energy
$E_{0}$ into Eqn.[\ref{weak_a}] and Eqn.[\ref{weak_b}], we
obtain the effective mass near the adiabatic limit as
\begin{equation}
\frac{m_{\mathrm{weak}}^{\ast }}{m_{SO}}=1+\frac{\lambda }{2}\frac{1}{\sqrt{%
1+\frac{(V_{R}+V_{D})^{2}}{2t^{2}}-\frac{(V_{R}-V_{D})^{2}}{%
(V_{R}+V_{D})^{2}}}}.
\end{equation}
The effective mass has a minimum for $V_{R}=V_{D}$ while $%
V_{R}+V_{D}$ is a constant.

\section{Strong coupling theory}

To investigate the strong coupling regime of the
Rashba-Dresselhaus-Holstein
model for a single polaron, we use the Lang-Firsov\cite{lang63} \cite%
{marsiglio95} unitary transformation $\overline{H}=e^{S}He^{-S}$ , where $%
S=g\sum_{i,\sigma }n_{i,\sigma }(a_{i}-a_{i}^{\dagger })$. Following procedures
similar to those in Ref. (\onlinecite{li11}), we obtain the
first order perturbation correction to the energy as
\begin{equation}
E_{k\pm }^{(1)}=e^{-g^{2}}\varepsilon _{k\pm }-g^{2}\omega _{E},
\end{equation}%
where $g$ is the band narrowing factor, as used in the Holstein
model. To find the second order correction to the ground state
energy, we proceed as in Ref. (\onlinecite{li11}), and find
\begin{eqnarray}
E_{k-}^{(2)} &=&-4e^{-2g^{2}}\frac{t^{2}+(V_{R})^{2}+\left(
V_{D}\right) ^{2}}{\omega _{E}}  \notag  \label{2nd_order_b} \\
&&\times
\bigl[f(2g^{2})-f(g^{2})\bigr]-e^{-2g^{2}}f(g^{2})\frac{\epsilon
_{k-}^{2}}{\omega _{E}},
\end{eqnarray}%
where $f(x)\equiv \sum\limits_{n=1}^{\infty }\frac{1}{n}\frac{x^{n}}{n!}%
\approx e^{x}/x\bigl[1+1/x+2/x^{2}+...\bigr]$. Thus the ground state
energy, excluding exponentially suppressed corrections, is
\begin{equation}
E_{GS}=-2\pi t\lambda \bigl(1+2\frac{t^{2}+(V_{R})^{2}+\left(
V_{D}\right) ^{2}}{(2\pi t\lambda )^{2}}\bigr),
\end{equation}%
and there is a correction of order $1/\lambda ^{2}$ compared to the
zeroth order result. Corrections in the dispersion enter in strong
coupling only with an exponential suppression. The ground state
energy predicted by strong
coupling theory has a maximum for $V_{R}=V_{D}$ while $%
V_{R}+V_{D}$ is a constant.

\end{document}